\documentstyle[epsfig]{aipproc}

\begin{document}
\title{Ionization Structure and the Reverse Shock in E0102-72}

\author{K.A.~Flanagan, C.R.~Canizares, D.S.~Davis, D.~Dewey, J.C.~Houck,
M.L.~Schattenburg}
\address{Center for Space Research, Massachusetts Institute of Technology\\
Cambridge, MA 02139}

%\lefthead{LEFT head}
%\righthead{RIGHT head}
\maketitle
\begin{abstract}
	The young oxygen-rich supernova remnant E0102-72 in the
Small Magellanic Cloud has been observed with the High Energy Transmission 
Grating Spectrometer of Chandra. The high resolution X-ray spectrum 
reveals images of the remnant in the light of individual emission
lines of oxygen, neon, magnesium and silicon. The peak emission
region for hydrogen-like ions lies at larger radial distance
from the SNR center than the corresponding helium-like ions, suggesting
passage of the ejecta through the "reverse shock". We examine models
which test this interpretation, and we discuss the implications.
\end{abstract}
\section*{Background}
1E0102.2-7219 (hereinafter referred to as E0102-72) 
is a young ($\sim$1000 years) oxygen-rich supernova remnant (SNR) 
in the Small Magellanic Cloud. 
%It was discovered in X-rays 
%with the {\it Einstein} Observatory (Seward and Mitchell, 1981).
%Optical filaments of oxygen were found 
%(Dopita {\it et al.}, 1981) and measured to have velocities of 
%thousands of km/s (Tuohy and Dopita,1983), identifying it as 
%an oxygen-rich SNRs.  Only a small number of oxygen-rich SNRs 
%have been identified: They are believed to come from 
%massive progenitors, exhibit high velocities in their optical filaments, 
%are young (with the possible exception of Puppis A) and often show 
%evidence of asymmetrical explosion. 
E0102-72 has been observed at all wavelengths: in the 
optical\cite{dopita81}\cite{tuohy83}, UV\cite{blair89}\cite{blair2000} 
, radio\cite{amy93}, and in the 
X-ray\cite{seward81}\cite{hayashi94}\cite{gaetz2000}\cite{hughes88}\cite{hughes2000}\cite{davis2000}\cite{canizares2000}\cite{rasmussen2000}. 
%Based on early X-ray observations, Hughes (1988) ruled out a 
%uniform-density shell model. 
%Our data (Houck, {\it et al.}, 2000) and Gaetz {\it et al.}
%confirm Hughes' conclusion. 
Gaetz {\it et al.} have looked at direct Chandra images of 
E0102-72 and noted the radial variation with 
energy bands centered on OVII and OVIII, suggesting an ionizing shock 
propagating inward. This paper addresses this issue through analysis
of the high resolution dispersed spectrum, and reinforces this interpretation.
%Gaetz {\it et al.} have looked at direct {\it Chandra} images of 
%E0102-72 and see the boundary of the
%blast-wave shock, inferring a blast wave velocity of $\sim$6200 km/sec.
%This estimate agrees well with Hughes {\it et al.}, 2000, who has measured
%proper motion in the supernova remnant and
%
\section*{The High Resolution X-Ray Spectrum}
Shown in Figure~1 is the dispersed high resolution X-ray spectrum of 
E0102-72.  The Chandra observation was made using the High Energy
Transmission Gratings (HETG) in conjunction with the Advanced 
CCD Imaging Spectrometer (ACIS-S). The HETG separates the 
X-rays into their distinct emission lines, forming separate 
images of the remnant with the X-ray light of each line. 
The dispersed spectrum contains lines of highly ionized oxygen, neon, 
magnesium and silicon (not shown in the figure). 
Detailed study of the line images shows they are distorted
along the dispersion axis due to Doppler shifts associated with high
velocity material. Details and an overview of results are given in 
Canizares {\it et al.}  elsewhere in these proceedings.  
Here we examine the apparent differences in ring radii
seen in Figure~1. As discussed in detail below, these are likely 
caused by the changing ionization state, 
suggesting the passage of the supernova ejecta through the reverse shock. 
%%%%%
\begin{figure}[h!] % fig 1
\centerline{\epsfig{file=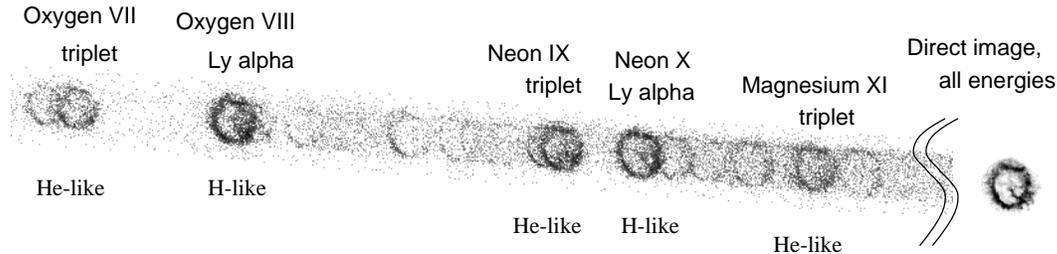,angle=-90,width=5.5in}}
\caption{A portion of the high resolution spectrum formed by the 
medium energy gratings.  At right in the figure is the zeroth order, 
which combines all energies in an undispersed image. }
\label{fig1}
\end{figure}
%\vspace{-.3in}
%
\begin{figure} %fig2
\begin{minipage}[h!]{.46\linewidth}
\hspace{-.25in}
\centering\epsfig{file=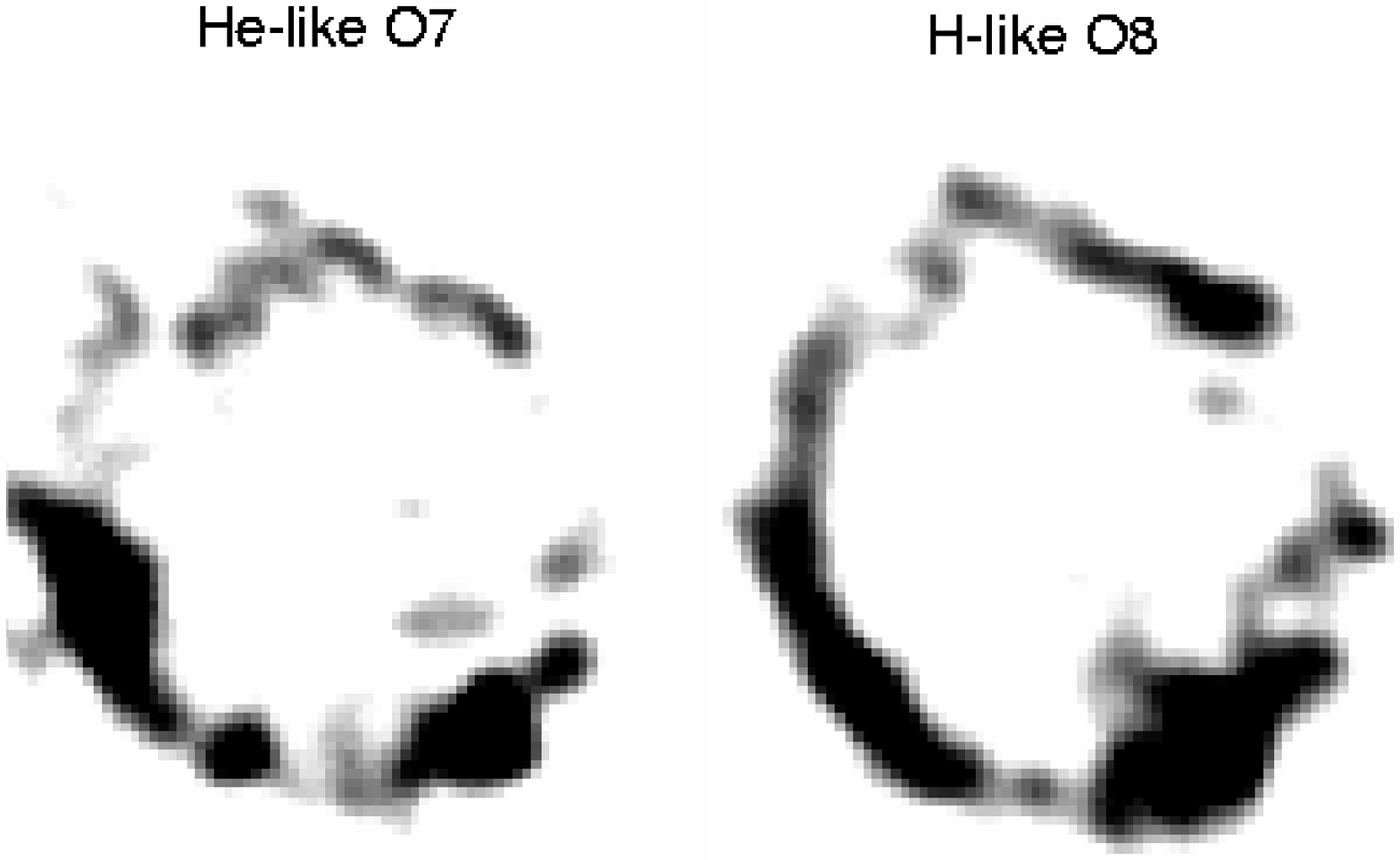, height=1.9in}
\caption{The OVII Resonance line is emitted from a region of smaller
radius than that of OVIII Lyman $\alpha$.}
\label{fig2}
\end{minipage}
\hfill
\begin{minipage}[h!]{.49\linewidth}
%\centering{\epsfig{file=manuscript_ox_diameter.eps,width=\linewidth}}
\centering{\epsfig{file=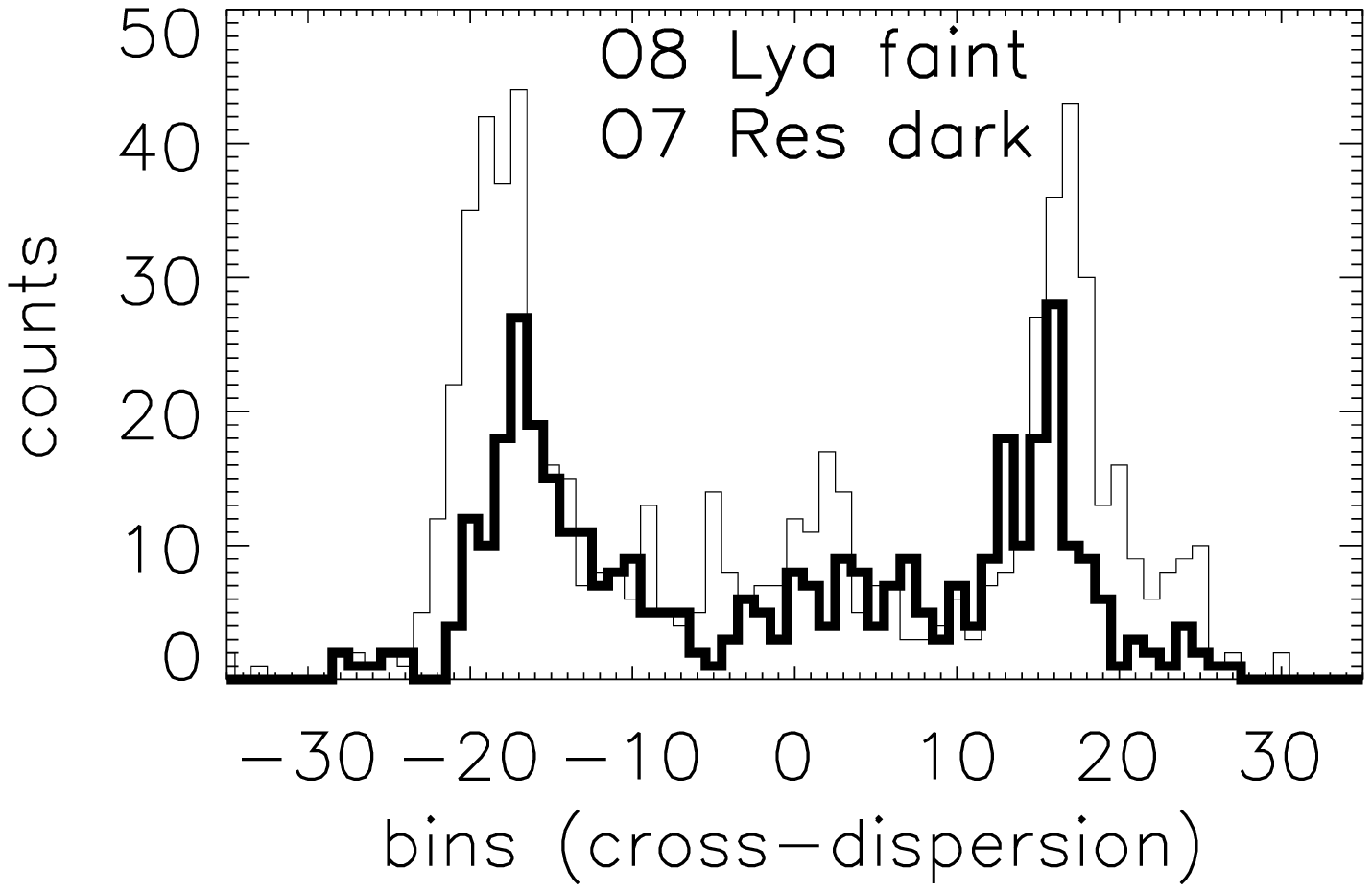,width=\linewidth}}
\caption{The OVII Resonance distribution is narrower than that
of OVIII~Lyman~$\alpha$.}
\label{fig3}
\end{minipage}
\end{figure}
\section*{The Ionization/Shock Structure of E0102-72}
Figure~2 compares helium-like and hydrogen-like lines of 
oxygen. The dispersed OVII Resonance line from the helium-like 
triplet (left) is displayed next to the hydrogen-like 
OVIII~Ly~$\alpha$ line (right) on the same spatial scale. 
The ring diameter of the hydrogen-like line is obviously 
larger than that of the helium-like line.
The ring diameters of these lines were measured by
tracing the distribution in the direction orthogonal to the dispersion
axis, as shown in Figure~3. (By taking the distribution in the 
cross-dispersion direction, the result is largely independent of 
Doppler shift.)  The He-like OVII distribution is
nestled cleanly within the OVIII, and is measurably narrower in Figure~3.
The ring diameters for all the bright lines were measured similarly,
and the results are shown on Figure~4. 
{\it All of the hydrogen-like lines
lie outside the corresponding helium-like lines.}
\vspace{-.15in}
\begin{figure}[h] % fig 4 height=2.1
\begin{minipage}{.46\linewidth}
%\centering{\epsfig{file=pic_ring_energy.ps,width=\linewidth}}
\centering{\epsfig{file=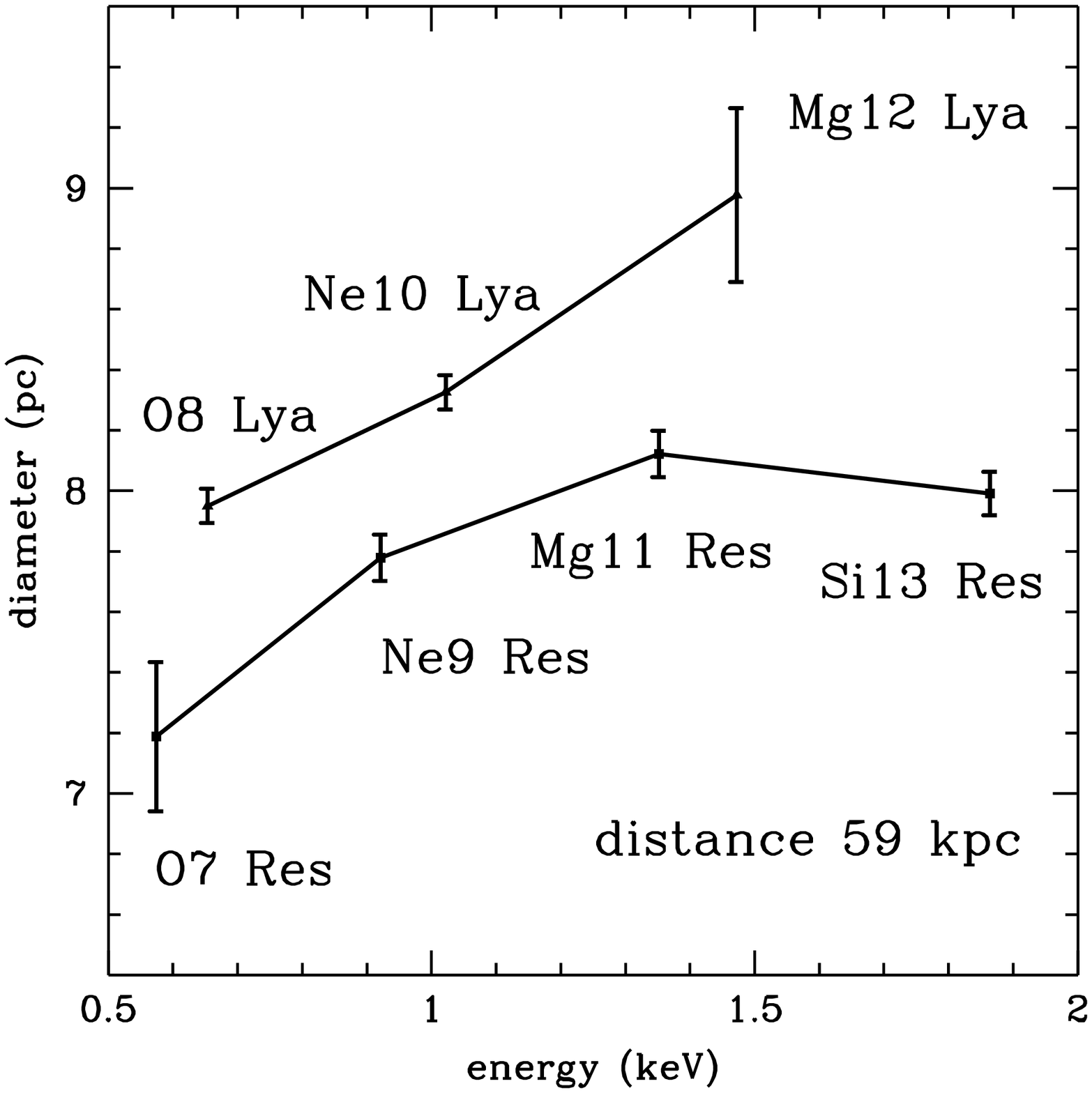,width=\linewidth}}
\caption{Ring diameter vs energy for the bright lines.  Note that
all the hydrogen-like lines (connected by the top curve)
lie outside their helium-like counterparts.}
\label{fig4}
\end{minipage}
\hfill
\begin{minipage}{.46\linewidth} % Fig 5 height=2.1 
%\centering{\epsfig{file=pic_emiss.ps,width=\linewidth}}
\centering{\epsfig{file=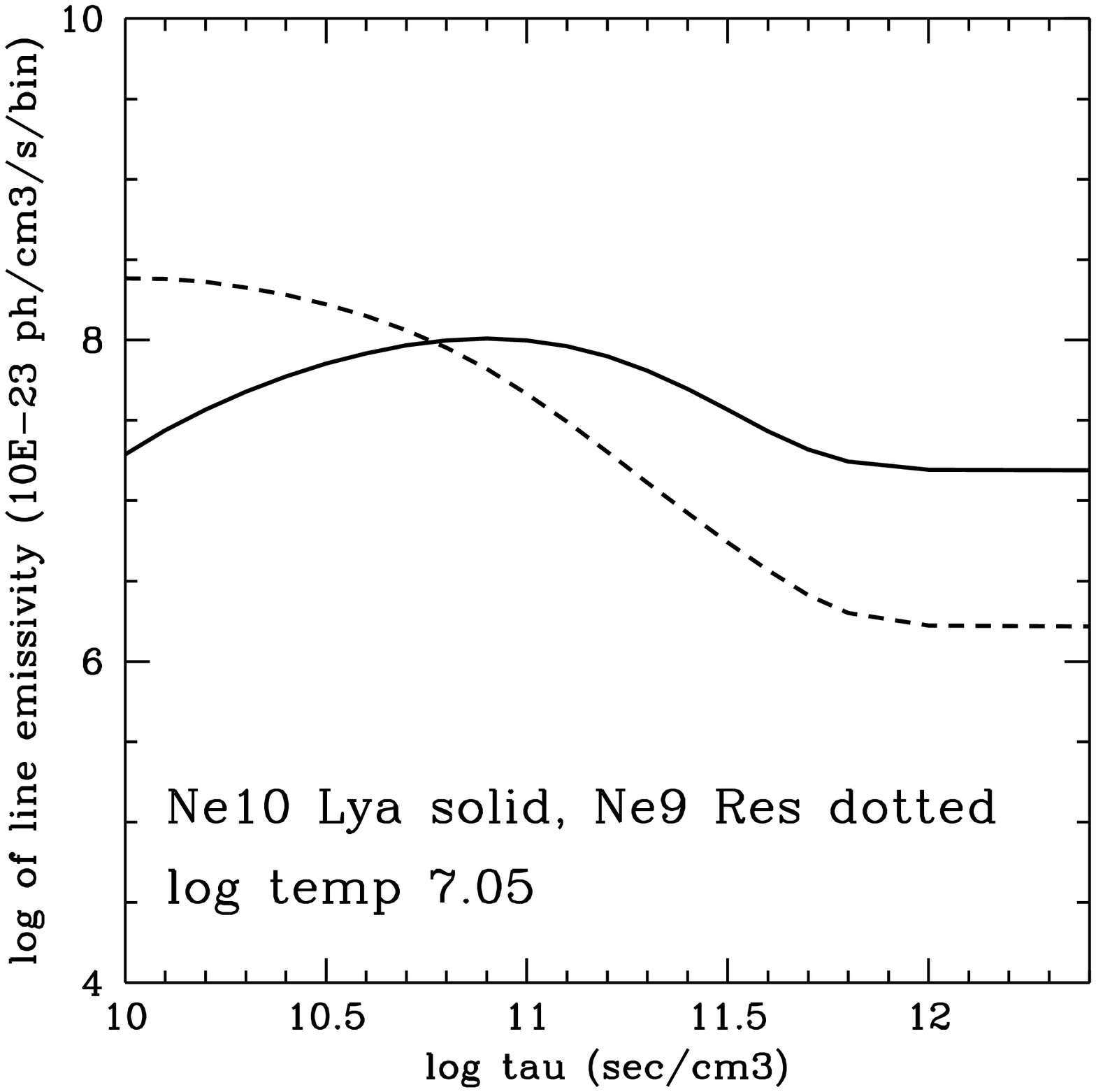,width=\linewidth}}
\caption{Helium-like NeIX~Resonance line reaches its peak emissivity at
log$\tau$=10.0, earlier than hydrogen-like NeX~Lyman~$\alpha$, which peaks at
about log~$\tau$=11.0 sec/cm$^3$}
\label{fig5}
\end{minipage}
\end{figure}
After passage of a shock, an ionizing plasma at a fixed electron 
temperature T$_e$ will achieve a He-like state before it
reaches a H-like state. This is shown in Figure~5, where
the appropriate timescale is the ionization parameter $\tau$=n$_e$t,
where t is the time since passage of the shock and n$_e$ is the electron
density\cite{hughes85}.  Since the H-like state lies outward of the He-like
state in E0102-72, this suggests the action of a shock moving {\it inward} 
relative to the ejecta - the `reverse shock' which is the standard model
for the mechanism which heats SNR ejecta to X-ray temperatures. 
\vspace{-.15in}
\begin{figure}[h!] 
\begin{minipage}[h]{.46\linewidth} % Fig 6
%\centering{\epsfig{file=pic_ion_705.ps, width=\linewidth}}
\centering{\epsfig{file=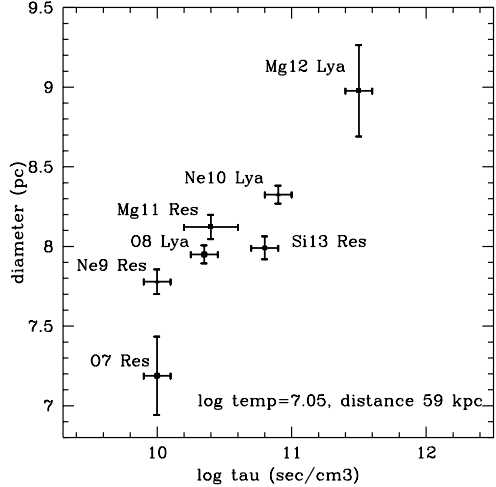, width=\linewidth}}
\caption{Measured diameter vs $\tau$ at peak emissivity. O7 is an upper 
limit for $\tau$.}
\label{fig6}
\end{minipage}
\hfill
\begin{minipage}{.46\linewidth}
We applied a simple model 
in which a homogenous mix of ejecta is
shocked to a fixed T$_e$ of about 1~keV or log T$_e$=10$^{7.05}$ 
(as suggested by our global NEI analysis\cite{davis2000}). 
We estimated $\tau$ by finding the peak emissivity (as in Figure~5).
A plot of measured diameters as a function of $\tau$ is shown in
Figure~6.  Although any workable model must consider many parameters,
the monotonic behavior suggests that these differences
in ring diameter are attributable to the ionization structure resulting
from the reverse shock.
\end{minipage}
\end{figure}
\section*{Work in Progress and Future Work}
 Our current model calls for further work.
%One could consider a plasma in ionization equilibrium at every point,
%with a radial temperature variation, however our plasma 
%diagnostics\cite{davis2000} consistently indicate a 
%nonequilibrium plasma. We may consider taking potentially useful plasma 
%diagnostic line ratios from the cross-dispersion histograms. 
For example, by taking the ratio of the histograms for 
O8~Lyman~$\alpha$ and O7~Resonance from Figure~3, we can examine the allowed 
region of paramater space defined by T$_e$ and $\tau$ {\it as a function of 
radial position}. (This approach as it pertains to the global spectrum 
is described by Davis {\it et al}\cite{davis2000} elsewhere in 
these proceedings.)
%Our global plasma disgnosticsIf the T$_e$ can be sufficiently constrained
%(i.e., to the ``asymptotic'' region where the ionization age is nearly
%independent of temperature), then we could examine the ionization structure 
%in detail. 
By considering a likely density gradient in the ejecta and
obtaining an appropriate estimate for electron 
density, we might obtain a corresponding estimate for the reverse 
shock velocity. We plan to apply this 
shock velocity information to see how it fits with the 
spatial extent of the bright ejecta and the presumed age of the remnant.
%%
%%\section*{acknowledgements}

\vspace{0.2in}

{\bf Acknowledgements}  We thank Glenn Allen, Norbert Schulz 
and Sara-Anne Taylor for helpful discussions. We are grateful 
to the CXC group at MIT for their advice and assistance. This work was 
prepared under NASA contract NAS8-38249 and SAO SV1-61010. 


\begin{references}
%
\bibitem{dopita81}Dopita, M.A., Tuohy, I.R. \& Mathewson, D.S., 1981, 
{\it ApJ}, {\bf 248}, L105 (1981)

\bibitem{tuohy83}Tuohy, I.R. \& Dopita, M.A.,{\it ApJ},{\bf 268}, L11 (1983)

\bibitem{blair89}Blair,~W.P., Raymond,~J.C., Danziger,~J. \& 
Matteucci,~F., {\it ApJ}, {\bf 338}, 812 (1989)

\bibitem{blair2000}Blair,~W.P., Morse,~J.A., Raymond,~J.C., Kirshner,~R.P., 
Hughes,~J.P., Dopita,~M.A., Sutherland,~R.S., Long,~K.S. \& Winkler,~P.F., 
{\it ApJ}, {\bf 537}, 667 (2000)

\bibitem{amy93}Amy, S.W. \& Ball, L., {\it ApJ}, {\bf 411}, 761 (1993)

\bibitem{seward81}Seward, F.D. \& Mitchel, M., {\it ApJ}, {\bf 243}, 736 (1981)
 
\bibitem{hayashi94}Hayashi,~I., Koyama,~K., Masanobu,~O., Miyata,~E., 
Tsunemi,~H., Hughes,~J.P. \& Petre,~R., {\it PASJ}, {\bf 46}, L121 (1994)

\bibitem{gaetz2000}Gaetz,~T.J., Butt,~Y.M., Edgar, R.J., Eriksen,~K.A., 
Plucinsky,~P.P., Schlegel,~E.M. \& Smith,~R.K., {\it ApJ}, {\bf 534}, L47 (2000)

\bibitem{hughes88}Hughes,~J.P. 1988, in {\it Supernova Remnants and the 
Interstellar Medium}, ed. R.S.~Roger \& T.L.~Landecker 
Cambridge: Cambridge Univ. Press, p.~125

\bibitem{hughes2000}Hughes,~J.P., Rakowski,~C.E. \& Decourchelle, A. 
{\it ApJ}, (2000) in press

\bibitem{davis2000}Davis, D.S., Flanagan, K.A., Houck, J.C., 
Canizares, C.R, Allen, G.E., Schulz, N.S., Dewey, D. \& Schattenburg, M.L.,  
{\it these proceedings}.

\bibitem{canizares2000}Canizares,~C.R, Flanagan,~K.A., Davis,~D.S., Dewey,~D.,
\& Houck,~J.C., {\it these proceedings}.

\bibitem{rasmussen2000}Rasmussen,~A. {\it et al.}, {\it  these proceedings}.

%\bibitem{flanagan2000}Flanagan, K.A., {\it et al.} 2000, in preparation


%\bibitem{houck2000}Houck,~J.C., {\it et al.} 2000, in preparation 

\bibitem{hughes85}Hughes, J.P. \& Helfand, D.J., {\it ApJ}, 
{\bf 291}, 544 (1985)

%\bibitem{raymond77}Raymond, J.C. \& Smith, B.W.,{\it ApJS},{\bf 35}, 419 (1977)

\end{references}
\end{document}